\documentclass[%
 reprint,
 amsmath,amssymb,
 aps,
]{revtex4-2}

\usepackage{graphicx}
\usepackage{dcolumn}
\usepackage{bm}

\usepackage{graphicx}
\usepackage{subcaption}
\usepackage{epsfig}
\usepackage{amssymb}
\usepackage{amsmath}
\usepackage[utf8]{inputenc}
\usepackage{graphicx}
\usepackage{xcolor}
\usepackage{float}
\newcommand{\s}{\langle s \rangle}

\begin{document}

\preprint{APS/123-QED}

\title{Dimensional analysis for clogging of grains in two and three dimensions}

\author{Julián Montero}
\affiliation{%
 Departamento de F\'isica, Facultad de Ciencias Exactas y Naturales,\\ Universidad Nacional de La Pampa, CONICET,\\ Uruguay 151, 6300 Santa Rosa (La Pampa), Argentina
}%

\author{Luis A. Pugnaloni}%
\email{luis.pugnaloni@exactas.unlpam.edu.ar}
\affiliation{%
 Departamento de F\'isica, Facultad de Ciencias Exactas y Naturales,\\ Universidad Nacional de La Pampa, CONICET,\\ Uruguay 151, 6300 Santa Rosa (La Pampa), Argentina
}%

\author{Ryan Kozlowski}
\affiliation{%
Physics Department, College of the Holy Cross, Worcester,
MA 01610, USA}%

\date{\today}

\begin{abstract}
We conduct standard dimensional analysis (Vaschy--Buckingham $\Pi$-theorem) for the mean avalanche size $\s$ when particles flow through, and clog at, a small orifice on the base of a flat-bottomed silo. We consider the effect of particle diameter $d$, orifice diameter $D$, particle density $\rho$, particle Young's modulus $E$ and acceleration of gravity $g$. We both perform discrete element method simulations and compile available data in the literature in order to sample the parameter space. We find that our simulations and data across many experiments and simulations of frictional grains are consistent with the scaling equation  $\ln (\s+1) = A_\alpha (D/d-1)^{\alpha} + B_\alpha \sqrt{\rho g d / E}$, where $A_\alpha$ and $B_\alpha$ are empirical constants and $\alpha$ is the dimensionality of the system ($\alpha=2$ and $\alpha=3$ for 2D and 3D, respectively). This expression successfully synthesizes the clogging behavior of a number of related clogging systems and motivates future extensions to more complex configurations involving, for example, very low friction particles or external vibrations. 
\end{abstract}

\keywords{Clogging \sep Granular flow }
                              
\maketitle


\section{Introduction}
Granular materials, composed of discrete solid particles, are prevalent in numerous natural and industrial processes. Their behavior is complex and exhibits the characteristics of solids, liquids, and gases. This complexity arises from the dissipative nature of interactions between grains. Understanding granular materials is crucial in various fields, including agriculture, mining and pharmaceuticals \cite{Duran2000}.

The flow of granular materials through constrictions, such as those found in the orifice of silos, is a fundamental area of study. Although the discharge rate is well defined for large constrictions, typically larger than five times the particle size if particles are spherical and cohesionless, the flow can be interrupted for small orifices by the formation of structures in the form of arches (2D) or vaults (3D) that can block the exit and stop the discharge \cite{Nedderman1992}. This flow blockage is a significant issue in many applications, causing disruptions in industrial processes and impeding the transport of materials. This phenomenon arises from the complex interplay of forces and geometric constraints as particles converge towards a narrow opening. Clogging is not limited to granular materials, but is observed in various systems involving the flow of discrete entities through bottlenecks, including colloids, sheep herds, and even pedestrian crowds \cite{Zuriguel2014}. In this context, Zuriguel (Ref.~\cite{Zuriguel_PIP_2014}) reviews the obstruction phenomenon, discussing statistical distributions in clogging/unclogging and the influence of particle shape and outlet size.

Researchers have explored various factors that influence granular flow and clogging. Among these, the geometry of the system plays a crucial role. This includes the shape and size of the aperture and the tilt of the hopper; for example, an increase in the angle of the hopper reduces the probability of clogging \cite{Lopez_Rodriguez_PRE_2019}. Beyond static geometric factors, external perturbations can also significantly impact granular flow. Such perturbations often lead to intermittent flow patterns, characterized by alternating periods of flow and arrest, with statistical distributions of clogging times playing a key role in understanding these underlying processes. Studies on vibration, for example, have shown that while horizontal oscillation of the base of a silo can help maintain flow for small exits, excessive amplitude oscillations at larger exits can paradoxically impede flow \cite{To}. Furthermore, research has explored how vibrations internal to particles, influencing their contact mechanics, can help prevent jamming and clogging, a hypothesis supported by recent simulations \cite{Parisi2023}.

The properties of the granular material -- be it intrinsic properties of the material composing grains or composite features of grain mixtures -- also influence flow properties and clogging statistics. For example, the introduction of a second, finer-grained species to an initially large, monosized granular material can increase the flow of the original species and reduce clogging occurrences. This counterintuitive finding, observed both in three-dimensional experimental studies \cite{Gharat_PowTech_2023} and in two-dimensional simulations \cite{Madrid2021}, has significant implications for optimizing industrial processes. A variety of studies also highlight the critical role of particle deformability on clogging, particularly in distinguishing between ``soft" (deformable) and ``hard" (rigid) particles. Deformable particles, such as hydrogels or emulsion droplets studied in a variety of experiments~\cite{Hong_PRE_2017, Tao_PRE_2021}, require much smaller hopper openings to clog compared to rigid grains. Complementary to this, Wang et al. highlight the unique effects of viscoelastic properties in granular flow, particularly when dealing with mixtures of soft and rigid grains \cite{Wang_Soft_Matter_2021}. Their work demonstrates that the combined elastic and viscous response of particles, along with frictional interactions, introduces qualitatively new features in flow characteristics and clogging behavior, notably when even small proportions of rigid grains are added to a soft ensemble~\cite{Alborzi2022,Alborzi2023,Alborzi2025}.

Despite these advances in understanding the mechanisms underlying granular clogging, a generalized physical formulation that connects the intrinsic properties of the system with a characteristic scale of the discharge phenomenon, such as the average size of avalanches preceding blockage, has yet to be found. It is precisely this fundamental gap that this work aims to address. Here, we conduct a standard dimensional analysis using the Vaschy-Buckingham $\Pi$-theorem to begin to explore this connection in the case of a non-vibrated, flat-base silo with monosized grains of varying stiffness. Our primary objective is to derive a scaling law that quantitatively describes the mean avalanche size ($⟨s⟩$) as a function of the most relevant physical parameters in this traditional silo configuration: the diameter of the particle ($d$), the diameter of the orifice ($D$), the particle density ($\rho$), the particle Young's modulus ($E$), and the acceleration due to gravity ($g$). We indeed find a scaling which collapses the results of numerous prior studies as well as our own simulations with widely varying grain stiffness. The success of this scaling suggests that such an analysis can be, in future studies, extended to more complex and realistic granular systems (e.g., mixtures of grain types, container geometries, perturbations, etc.).

In Section II, we use the Vaschy-Buckingham $\Pi$-theorem to pose a relevant set of nondimensional parameters and, based on available data from the literature, present a general scaling law. In Section III, we describe the simulations we used to test the proposed scaling law in regions of the parameter space not accessed in prior works. In Section IV, we present our simulation results for 2D and 3D configurations along with prior data from the literature, and in Section V we discuss the validity of our scaling law and motivate the further application of the $\Pi$-theorem to more generalized systems exhibiting clogging.

\section{$\Pi$-theorem for clogging systems}

The mathematical formalization of dimensional analysis is called the Vaschy--Buckingham $\Pi$-theorem and is usually attributed to Buckingham \cite{Buckingham1914} despite similar, earlier works by Vaschy \cite{Vaschy1892}, Bertrand \cite{Bertrand1879} and Riabouchinsky \cite{Riabouchinsky1911}. In brief, this theorem states that if there is an equation that relates $n$ physical quantities $Q_i$ and these quantities can be expressed in terms of $k$ basic physical units of measure (mass, length, time, etc.), due to the dimensional homogeneity of physical equations, this equation is equivalent (i.e., has the same set of solutions) to another that relates $n-k$ non-dimensional quantities $\Pi_i$ which are defined as products of $Q_i$ elevated to some convenient exponents ($\Pi=Q_1^{\beta_1}Q_2^{\beta_2}...Q_n^{\beta_n}$). There are several, equally valid choices for the $n-k$ parameters $\Pi_i$, which leaves room for selecting the one that suits to the physical interpretation of interest.        

Let us consider the mean avalanche size $\s$ that flows through an orifice before clogging. We can assume from previous experiment by various authors \cite{To2001,Zuriguel2005,Janda2008,To2005,Thomas2015,Zuriguel2003,Arevalo2016,Hong_PRE_2017,Tao_PRE_2021,Alborzi2025} that $\s$ depends on the grain size $d$, the orifice size $D$, the material density $\rho$, the material Young modulus $E$ and the acceleration of gravity $g$. There are thus six physical quantities involved, expressed in terms of three physical units (mass, length, and time). Therefore, there exists only three independent non-dimensional parameters $\Pi_1$,  $\Pi_2$ and $\Pi_3$ which can be related through a dimensionally homogeneous physical equation:

\begin{equation}
 \Pi_1 = f(\Pi_2,\Pi_3),
\end{equation}
with $f$ an unknown function.

The $\Pi$s can be selected at will. So we chose the following non-dimensional quantities: $\Pi_1=\s$, $\Pi_2=D/d$ and $\Pi_3=\rho g d / E$. Here, $\Pi_1$ is our quantity of interest, which is non-dimensional {\it per-se}. $\Pi_2$ can be interpreted as the ``effective size'' of the orifice in terms of the size of the particles \cite{To2001}. Finally, $\Pi_3$ measures the relation between the weight of a grain in gravity ($\propto \rho g d^3$) and the force needed to compress a material of Young's modulus $E$ with a given cross section ($\propto Ed^2$) and can therefore be considered as a measure of ``effective softness'' of the grains \cite{Hong_PRE_2017}. Hence,

\begin{equation}
 \s = f\left(\frac{D}{d},\frac{\rho g d}{E}\right).
\end{equation}

In the literature, a number of authors have observed through experiments \cite{To2005,Janda2008,Thomas2015} and using statistical models \cite{Thomas2015} that $\s \propto \exp[ (D/d)^{\alpha} ]$, with $\alpha=2$ in 2D and $\alpha=3$ in 3D. There is an exception to this functional form in which authors propose a divergence of $\s$ at finite $D/d$ \cite{Zuriguel2003,Zuriguel2005} for 3D systems. However, with the available data, it is hard to be conclusive as to whether such divergency exists since the fast increasing $\exp[ (D/d)^3]$ law is equally successful at fitting the data.

Some authors have run experiments with various materials (plastic, glass, lead, steel, etc.) and various particle sizes \cite{Zuriguel2005,Thomas2015}, often with grains on the order of $d\sim 1$~mm. These materials have large Young's modulus (10 GPa $<E<$ 200 GPa). Since $\rho g d$ is small (10 Pa $< \rho g d < 100$ Pa) for 1-mm particles of densities between that of water and lead, the effect on $\Pi_3=\rho g d/E$ is minute with $\Pi_3 \rightarrow 0$. We consider these experiments, therefore, to be operating in a hard-grain limit. Simulations studying clogging still tend to operate in the hard-grain limit despite the tendency to run them at lower Young's modulus ($E\sim 10$~MPa) for computational efficiency. 

Despite our limited knowledge of the role of particle stiffness for clogging statistics, there are a few studies to our knowledge that suggest a potential scaling exponent for $\Pi_3$. On the one hand, Arévalo et al. \cite{Arevalo2016} have run simulations with a softer particle interaction (linear spring instead of Hertzian) in which they vary $g$ by four orders of magnitude. In these simulations they observed that $\s \propto \exp[\sqrt{g}]$. On the other hand, Hong et al. \cite{Hong_PRE_2017} have run experiments with very soft oil droplet particles. Their data suggests $\s \propto \exp[ \sqrt{\rho g d / E}]$. These experiments and simulations were done in 2D or quasi-2D configurations. However, since these experiments were run in a hopper rather than a flat bottom silo -- an important change to the silo geometry that has been shown to influence clogging with hard particles~\cite{Lopez_Rodriguez_PRE_2019} --  it is difficult to directly compare across these different studies. Also, to our knowledge, no 3D experiments or simulations are available for the clogging of soft particles where the mean avalanche size can be extracted. Finally and crucially, the soft particles used in experiments are based on droplets \cite{Hong_PRE_2017,Hong2022} or hydrogel beads \cite{Hong2022,Tao_PRE_2021,Alborzi2022}. These grains, apart from being soft, present extremely low friction coefficients, which add a new parameter to the problem. Indeed, friction coefficient has been shown to influence clogging properties for coefficients below 0.1~\cite{Zablotsky2024,Alborzi2025}.

Based on the available results mentioned above, we propose that the following scaling should hold

\begin{equation}\label{eq:scaling}
 \ln(\s +1)= A_\alpha \left(\frac{D}{d}-1\right)^{\alpha} + B_\alpha \sqrt{\frac{\rho g d}{E}},
\end{equation}
with $A_\alpha$ and $B_\alpha$ fitting constants that may depend on dimension $\alpha$. It is worth mentioning here that Eq.~(\ref{eq:scaling}) assumes that the mean avalanche size drops to zero if $D/d=1$ (i.e., the particles are of the size of the orifice) and $\rho g d/E=0$ (i.e., the particles are infinitely hard). Negative $\s$ is, of course, nonphysical. For soft particles ($\rho g d/E>0$), finite values of $\s$ are possible in Eq.~(\ref{eq:scaling}) even if $D<d$. Certainly, soft particles under large pressures can squeeze through an orifice smaller than the particle's non-compressed diameter, but we do not explore this extreme limit in the current work.

In the rest of the paper, we address the missing gap of the dependence of $\langle s\rangle$ on $\Pi_3$ posed in Eq.~(\ref{eq:scaling}) by exploring the soft frictional particle limit, for $D>d$, using 2D and 3D simulations. 

\section{Simulations}

To obtain numerical data for the mean size of the avalanche $\langle s \rangle$ as a function of dimensionless parameters $\Pi_{2} = D/d$ and $\Pi_{3} = \rho g d/E$, we performed discrete element method (DEM) simulations in two- and three-dimensional configurations using the open source software LIGGGHTS (LAMMPS Improved for General Granular and Granular Heat Transfer Simulations) \cite{Kloss2012}.

DEM models the dynamics of individual particles by explicitly integrating Newton's equations of motion, accounting for both contact interactions and external forces. In our simulations, particle interactions are governed by the nonlinear Hertz contact model with history-dependent tangential forces. This model accounts for elastic deformation during contact and includes both normal and tangential damping, providing a realistic description of the forces that arise during grain-grain and grain-wall collisions. 

The material properties were consistent across simulations: Poisson's ratio for both particles and walls was fixed at 0.3, the coefficient of restitution for all interactions was 0.9, and the friction coefficient was 0.5. The Young's modulus of the walls was maintained at $10^8$ Pa in all configurations. For particles, Young's modulus was fixed at $10^8$ Pa in 2D simulations, while in 3D simulations it was systematically varied.

We implemented a simulation protocol designed to reproduce gravity-driven discharge in silos under steady-state conditions with particle reinjection. Each simulation begins with a filling stage, where spherical particles are inserted from above into the silo with the exit orifice closed. Once the desired number of particles is reached (approximately $2500$ in 2D and $16000$ in 3D), the insertion stops. The system is then allowed to relax until it reaches a mechanically stable state under gravity with negligible kinetic energy ($<10^{-9}$ J).

\begin{figure}
    \centering
    \includegraphics[width=1.0\linewidth]{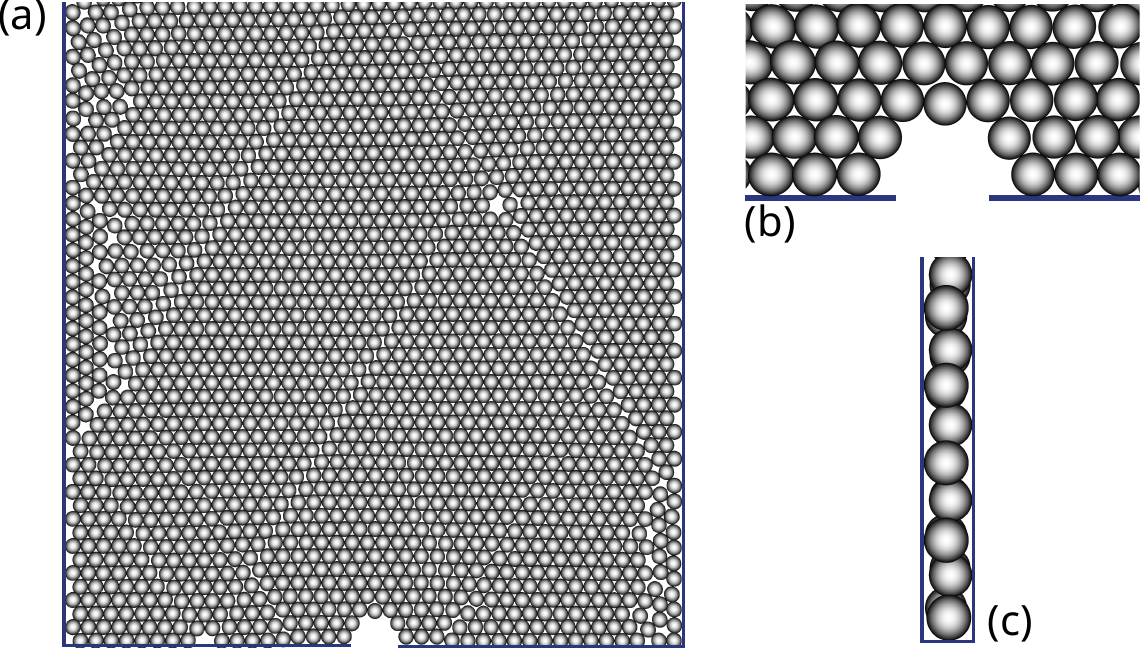}
    \caption{(a) Front view of a simulated quasi-2D system while clogged. (b) Closeup view of the clogging arch. (c) Side view of the silo showing the small game left for the particles in the flat prismatic container. The simulation was performed with $D/d =3.0$ and $E=10^8$ Pa.}
    \label{fig:2Dview}
\end{figure}

Once stabilized, the outlet opens to initiate discharge. The simulation domain has periodic boundaries in the vertical direction, ensuring particles exiting through the orifice are continuously reinserted from the top at their original horizontal positions. This enables collection of multiple avalanches per simulation run. An avalanche is measured by counting particles crossing a detection plane below the orifice. The avalanche size $s$ is defined as the number of grains flowing out between clogging arch breakage and reformation. Stable clogs are identified after waiting ten times the theoretical free-fall time of a particle through the orifice: $t_{\text{wait}} = 10 \sqrt{D/g}$. If no particles flow through the orifice during $t_{\text{wait}}$, we trigger a new avalanche. To this end, all particles within a region above the orifice are perturbed by setting their vertical velocity to  $v_z = 6d / t_{\text{pert}}$ and random horizontal velocities ($\pm 0.5v_z$) for a time interval of $t_{\text{pert}} = 0.5$~s. The perturbation region is cylindrical in 3D and rectangular in 2D ($20d$ in lateral width and $6d$ in height). This displaces particles vertically by $6d$ while random horizontal speeds prevent bias. After applying the perturbation, we wait for stabilization period defined as $t_{\text{relax}} = \sqrt{12d/g}$ before recording crossings through the orifice again. This procedure repeats until 200 avalanches are recorded, with $\s$, standard deviation, and standard error computed for statistical robustness.

In 2D, particles ($d=0.01$~m, $\rho=1500$~kg/m$^3$, $E=10^8$~Pa) were simulated in a flat prismatic silo (width $40.2d$, height $65.0d$, thickness $1.1d$). Representative images of the 2D system are shown in Fig.~\ref{fig:2Dview}. The orifice diameter $D$ is varied between $0.015$~m and $0.04$~m, which corresponds to $1.5<\Pi_{2}<4.0$. To vary $\Pi_{3}$, we adjust $g$ in the range $9.8$ to $980$ m/s$^{2}$, which corresponds to $1.47 \times 10^{-6}<\Pi_{3}<1.47 \times 10^{-4}$. We do not vary $E$ or $\rho$ in these simulations because there is already data available in the literature for different $E$ and $\rho$ in 2D that we will use to test the proposed scaling. 

\begin{figure}
    \centering
    \includegraphics[width=0.8\linewidth]{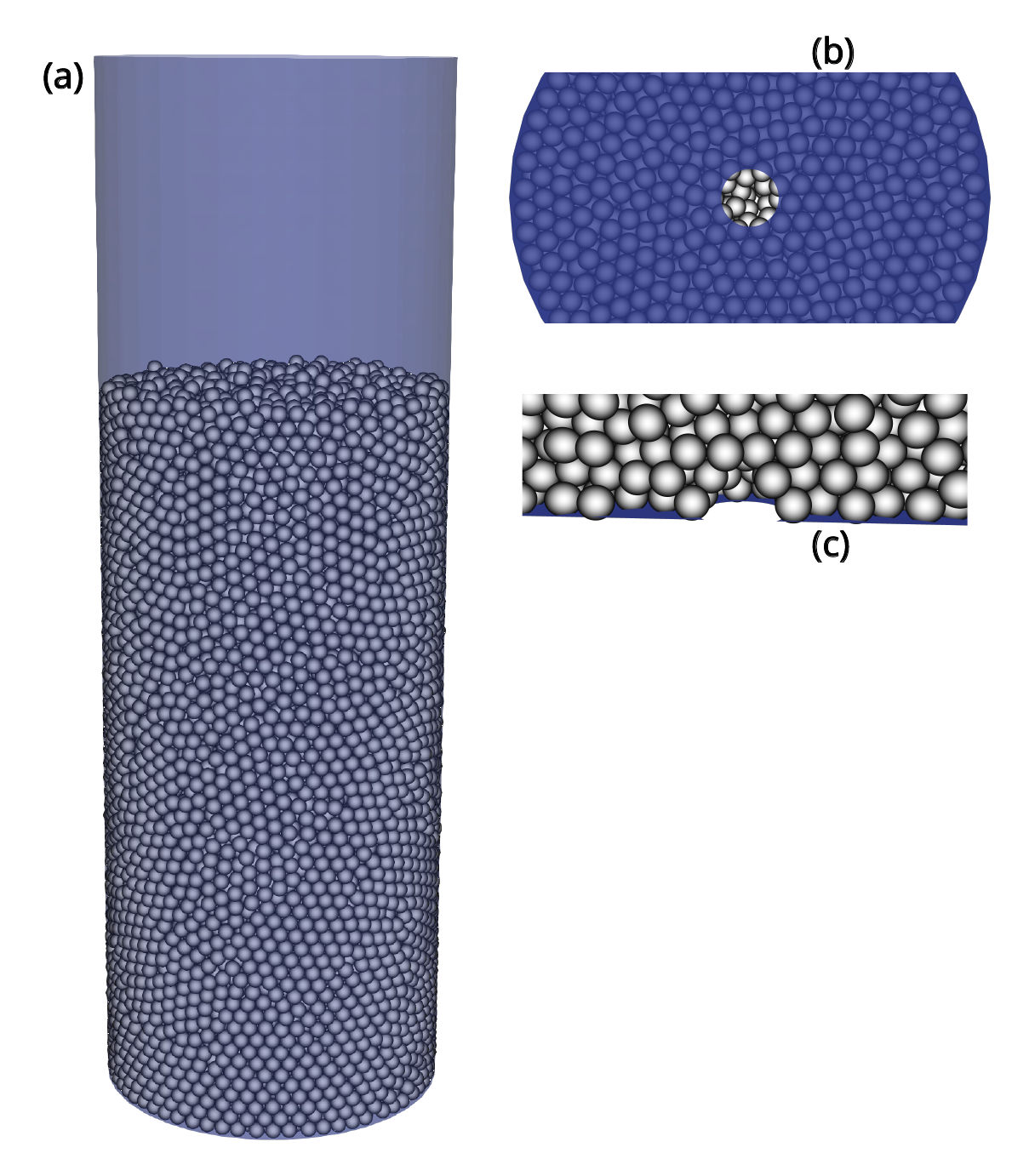}
    \caption{(a) 3D simulated system while clogged. (b) Bottom view of the clogged orifice. (c) Slice of the system around the orifice to show part of the clogging arch. This simulation was run with $D/d=2.6$ and $E=10^8$ Pa.}
    \label{fig:3Dview}
\end{figure}

In 3D, a cylindrical silo (diameter $21.7d$, height $65.2d$) contained particles of diameter $d=0.092$~m (see Fig.~\ref{fig:3Dview}). Orifice diameter was varied in the range $0.2024 < D< 0.368$~m, which corresponds to $2.0<\Pi_{2}<4.0$. There is not data in the literature for the clogging of soft particles in 3D configurations. Hence, to test Eq.~(\ref{eq:scaling}), we decided to assess the effect of $E$, $\rho$ and $g$ in our simulations. To vary $\Pi_{3}$ in the range $2.25 \times 10^{-6}<\Pi_3<6.24 \times 10^{-4}$, we explored the following ranges: $1000<\rho<4000$ kg/m$^{3}$; $1.3<g<271.39$ m/s$^2$; and $10^7<E<10^9$ Pa.

\section{Results}

\subsection{3D simulations}

\begin{figure}[t]
\centering
    (a)   \includegraphics[width=0.938\columnwidth]{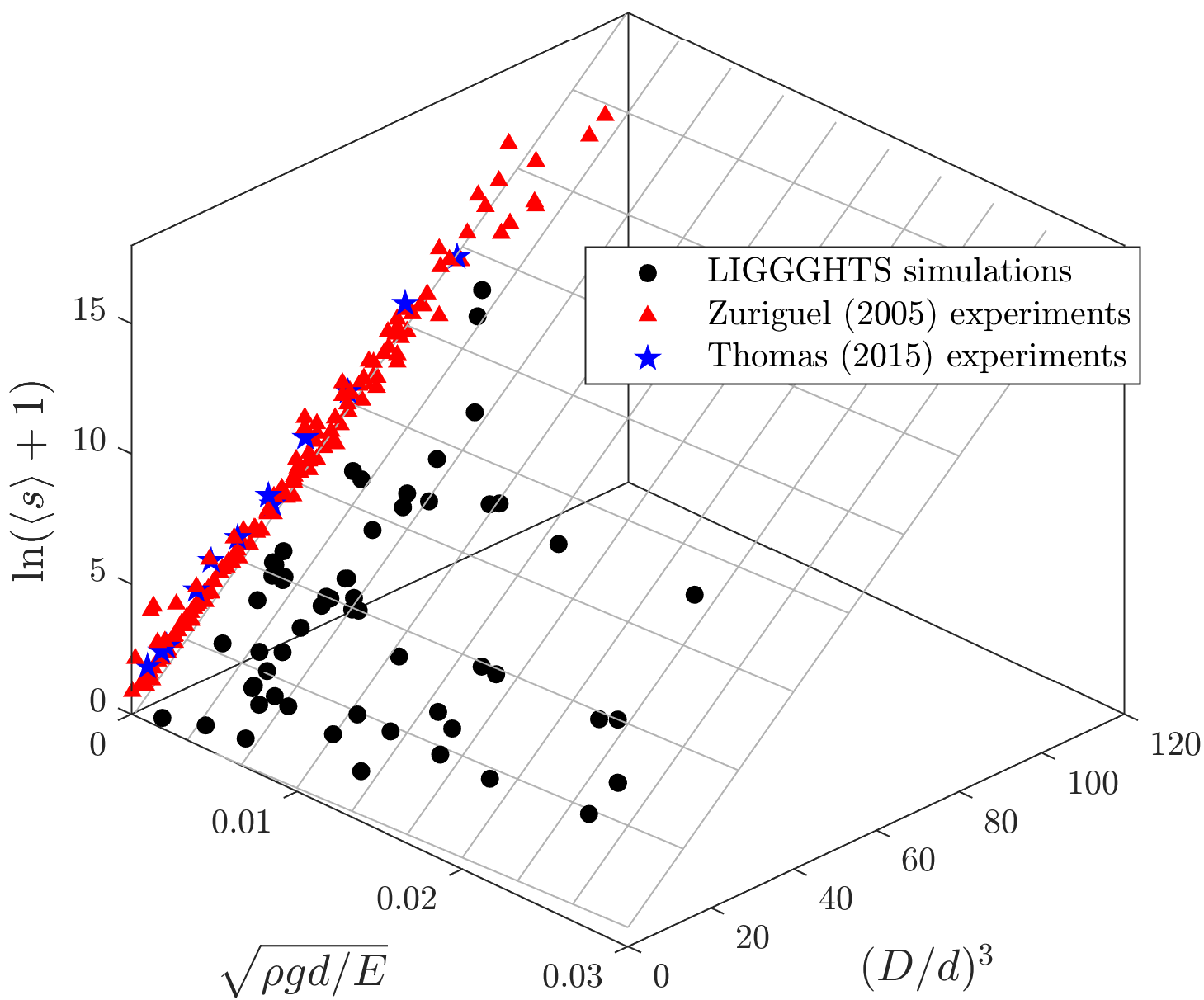}
    (b) \includegraphics[width=0.938\columnwidth]{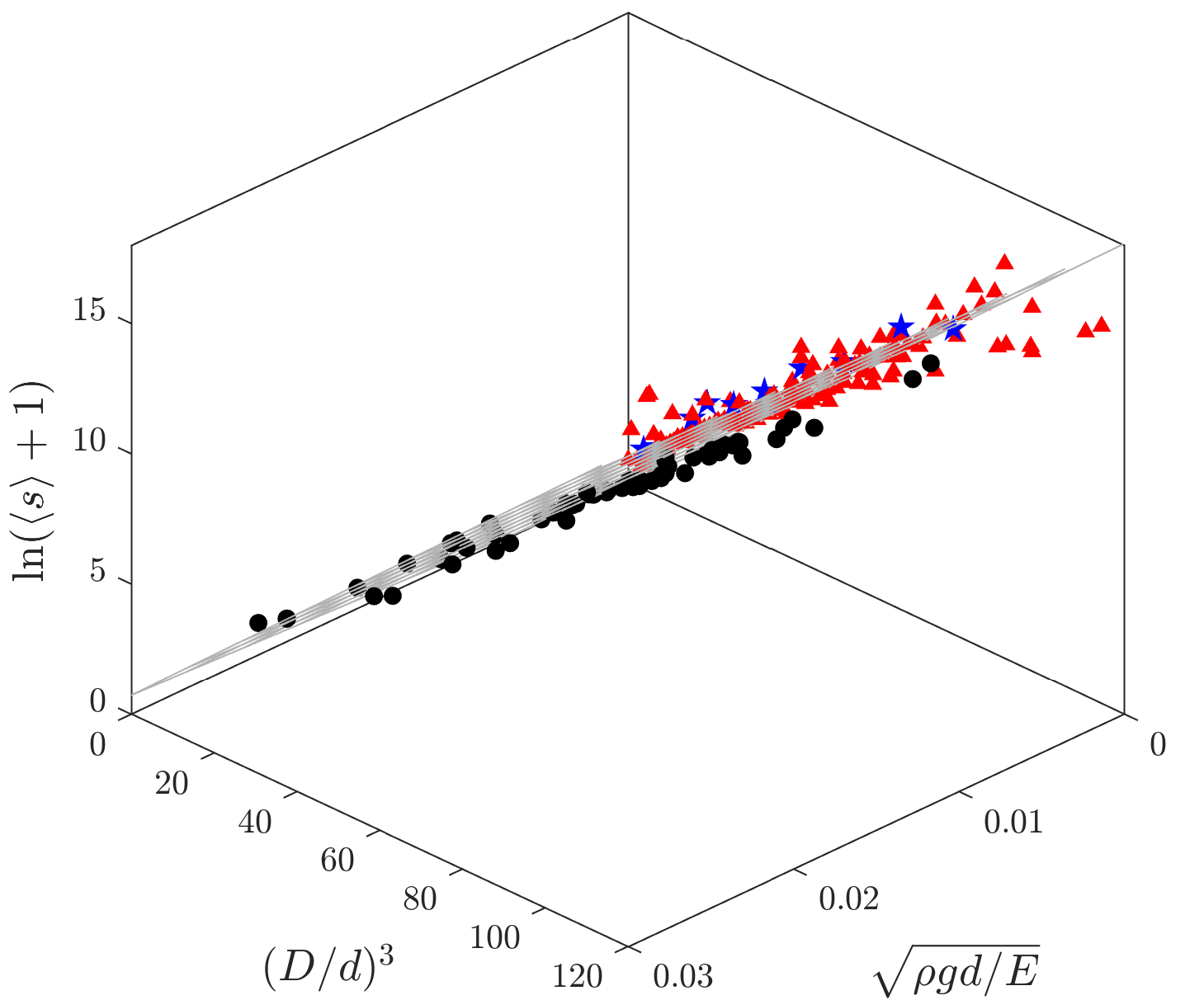}
    \caption{Mean avalanche size in a 3D silo as a function of the two dimensionless parameters. (a) and (b) show two different views of the 3D representation. Black dots represent our simulation data. Red triangles and blue stars correspond to the experimental values for hard grains from Refs. \cite{Zuriguel2005} and \cite{Thomas2015}, respectively. The plane fitting the data corresponds to $\ln(\s+1)=0.152 (\Pi_2-1)^3 + 29 \sqrt{\Pi_3}$. A rotating animation of this plot, as well as a Matlab .fig file, is available in the Supplemental Materials \cite{SM}.}
    \label{fig:3D}
\end{figure}

In Fig.~\ref{fig:3D}, we show $\ln(\s+1)$ as a function of both $\Pi_2^3=(D/d)^3$ and $\sqrt{\Pi_3}=\sqrt{\rho g d/E}$ from our simulations (black points). The parameter $\Pi_2$ was varied between $2.0$ and $4.0$ by adjusting $D$ while keeping $d$ constant. The  $\Pi_3$ parameter was varied in the range $10^{-6} \lesssim \Pi_3 \lesssim 10^{-3}$ by varying $\rho$, $g$ and $E$ as described in section III. The mean avalanche sizes in these simulations are in the range $4 \lesssim \s \lesssim 22000$. 

Along with the simulation results, we plot in  Fig.~\ref{fig:3D} experimental data extracted from Ref.~\cite{Zuriguel2005} (red triangles) and Ref.~\cite{Thomas2015} (blue stars). In Table \ref{tab:materials}, we summarize the material properties of all data extracted from the literature from these references as well as 2D studies we will present in Sec.~\ref{sec:2dsim}. These experiments were run in a configuration similar to our simulations with a flat bottom silo and glass beads, so we estimated $\rho=2500$ kg/m$^3$ and $E=7.0 \times 10^{10}$ Pa. Since the particle material is stiff, $\Pi_3$ is very small in these experiments. However, the experimental data is in line with our simulations for the higher Young modulus and lower acceleration of gravity simulated.   

\setlength{\tabcolsep}{10pt}
\begin{table*}[]
    \centering
    \begin{tabular}{c|rrrrcc}
    \hline\hline
        Material & d [m] &$\rho$ [kg/m$^3$] & $E$ [Pa] & $\mu$ & Ref. & Comments   \\
        \hline
        Glass spheres & 0.0025 & 2300 & *$7.0\times 10^{10}$ & 0.9 & Zuriguel (2005) \cite{Zuriguel2005} & 3D Flat Bottom\\
        Glass spheres & 0.0027 & 2500 & *$7.0\times 10^{10}$ & 0.9 & Zuriguel (2005) \cite{Zuriguel2005} & 3D Flat Bottom\\
        Glass spheres & 0.0020 & 2400 & *$7.0\times 10^{10}$ & 0.9 & Zuriguel (2005) \cite{Zuriguel2005} & 3D Flat Bottom\\
        Glass spheres & 0.0005 & 2540 & *$7.0\times 10^{10}$ & 0.9 & Zuriguel (2005) \cite{Zuriguel2005} & 3D Flat Bottom\\
        Glass spheres & 0.0020 & 2540 & *$7.0\times 10^{10}$ & 0.9 & Zuriguel (2005) \cite{Zuriguel2005} & 3D Flat Bottom\\
        Lead shots & 0.0025 & 10900 & *$1.4\times 10^{10}$ & 0.9 & Zuriguel (2005) \cite{Zuriguel2005} & 3D Flat Bottom \\
        Delrin beads & 0.0030 & 1340 & *$3.5\times 10^{9}$ & 0.37 & Zuriguel (2005) \cite{Zuriguel2005} & 3D Flat Bottom \\
        Steel beads & 0.0010 & 7600 & *$2.0\times 10^{11}$ & 0.7 & Zuriguel (2005) \cite{Zuriguel2005} & 3D Flat Bottom \\
        Glass spheres & 0.0020 & 1620 & *$7.0\times 10^{10}$ & 0.9 & Thomas (2015) \cite{Thomas2015} & 3D Flat Bottom \\
        Steel beads & 0.0010 & *7850 & *$2.0\times 10^{11}$ & 0.7 & Janda (2008) \cite{Janda2008} & 2D Flat Bottom\\
        Steel disk & 0.0050 & *7850 & *$2.0\times 10^{11}$ & 0.7 & To (2001) \cite{To2001} & 2D Flat Bottom\\
        Hydrogel beads & 0.0079 & 1030 & $3.5\times 10^{4}$ & 0.04 & Alborzi (2022) \cite{Alborzi2022} & 2D Hopper 34°\\
        Hydrogel beads & 0.0095	& 1030 & $3.5\times 10^{4}$ & 0.04 & Alborzi (2022) \cite{Alborzi2022} & 2D Hopper 34°\\
        Hydrogel beads & 0.0011 & 1030 & $3.5\times 10^{4}$ & 0.04 & Alborzi (2022) \cite{Alborzi2022} & 2D Hopper 34°\\
        Hydrogel beads & 0.0095 & 1030 & $3.5\times 10^{4}$ & 0.04 & Alborzi (2023) \cite{Alborzi2023} & 2D Hopper 34°\\
        Hydrogel beads & 0.0011 & 1030 & $3.5\times 10^{4}$ & 0.04 & Alborzi (2023) \cite{Alborzi2023} & 2D Hopper 34°\\
        Hydrogel beads & 0.0095 & 1030 & $3.5\times 10^{4}$ & 0.04 & Alborzi (2025) \cite{Alborzi2025} & 2D Hopper 34°\\
        Hydrogel beads & 0.0138 & *1000 & $5.4\times 10^{4}$ & 0.004 & Tao (2021) \cite{Tao_PRE_2021} & 2D Hopper 34°\\
        Glass beads & 0.0155 & 2500 & $8.0 \times 10^{10}$ & 0.9 & Tao (2021) \cite{Tao_PRE_2021} & 2D Hopper 34°\\
        Hydrogel beads & 0.013 & *1000 & $1.4\times 10^{5}$ & 0.006 & Hong (2017) \cite{Hong_PRE_2017} & 2D Hopper 34°\\
        Simulation & 0.001 & *2500 & **$2.45 \times 10^{6}$ & 0.5 & Arevalo (2015) \cite{Arevalo2016} & 2D Flat Bottom\\ 
        Simulation & 0.005 & 800 & *$1.0\times 10^{5}$ & 0.5 & Pérez (2000) \cite{Perez2008} & 2D Flat Bottom\\
        Simulation & 0.01 & 0.01 & *$\infty$ & 0.5 & Goldberg (2018) \cite{Goldberg2018} & 2D Flat Bottom\\
        \hline\hline
    \end{tabular}
    \caption{Summary of material properties extracted from the literature on clogging. Values marked with * where estimated from tables, not reported by the original authors; the value marked with ** (for Ref. \cite{Arevalo2016}) was estimated from a linear spring contact model, as described in Sec.~\ref{sec:2dsim}. \label{tab:materials} }
\end{table*}

We fit equation Eq.~(\ref{eq:scaling}) to the data in Fig.~\ref{fig:3D} and show the corresponding plane. The fitting constants are $A_2=0.152 \pm 0.002$ and $B_2 = 29 \pm 10$. As we can observe in Fig.~\ref{fig:3D}, the alignment of the data with the proposed scaling is fair despite the somewhat large dispersion that leads to the large error in $B_2$. It is important to bear in mind that mean avalanche size is affected to some extent by the protocol used to break arches \cite{Goldberg2018}. This means that the agreement between different labs and between experiments and simulations cannot be expected to be perfect since the method used to reinitiate flow after a clog is not always precisely described or quantified, and reproducing this aspect of the experiments is difficult. 

Unfortunately, there is no data available in the literature for large $\Pi_3$. Although experiments in 3D silos with soft particles have been carried our in recent years (see for example Refs.~\cite{Pongo2021,Stannarius2019}), these do not report clogging probability or mean avalanche size. It would be interesting to see if future experiments confirm the scaling in 3D for soft grains.

\subsection{2D simulations}\label{sec:2dsim}

Clogging probability and mean avalanche size have been studied more extensively in two-dimensional and quasi-two-dimensional configurations. Therefore, there is much more data available in the literature in this case. We have run simulations varying the non-dimensional parameters in the ranges $1.5 \lesssim D/d \lesssim 4.0$ and $10^{-6} \lesssim \rho g d/E \lesssim 10^{-4}$. To this end, we varied only $D$ and $g$ while we fixed $d$, $\rho$ and $E$, as described in section III.

\begin{figure}[t]
    \centering
    (a) \includegraphics[width=0.9\columnwidth]{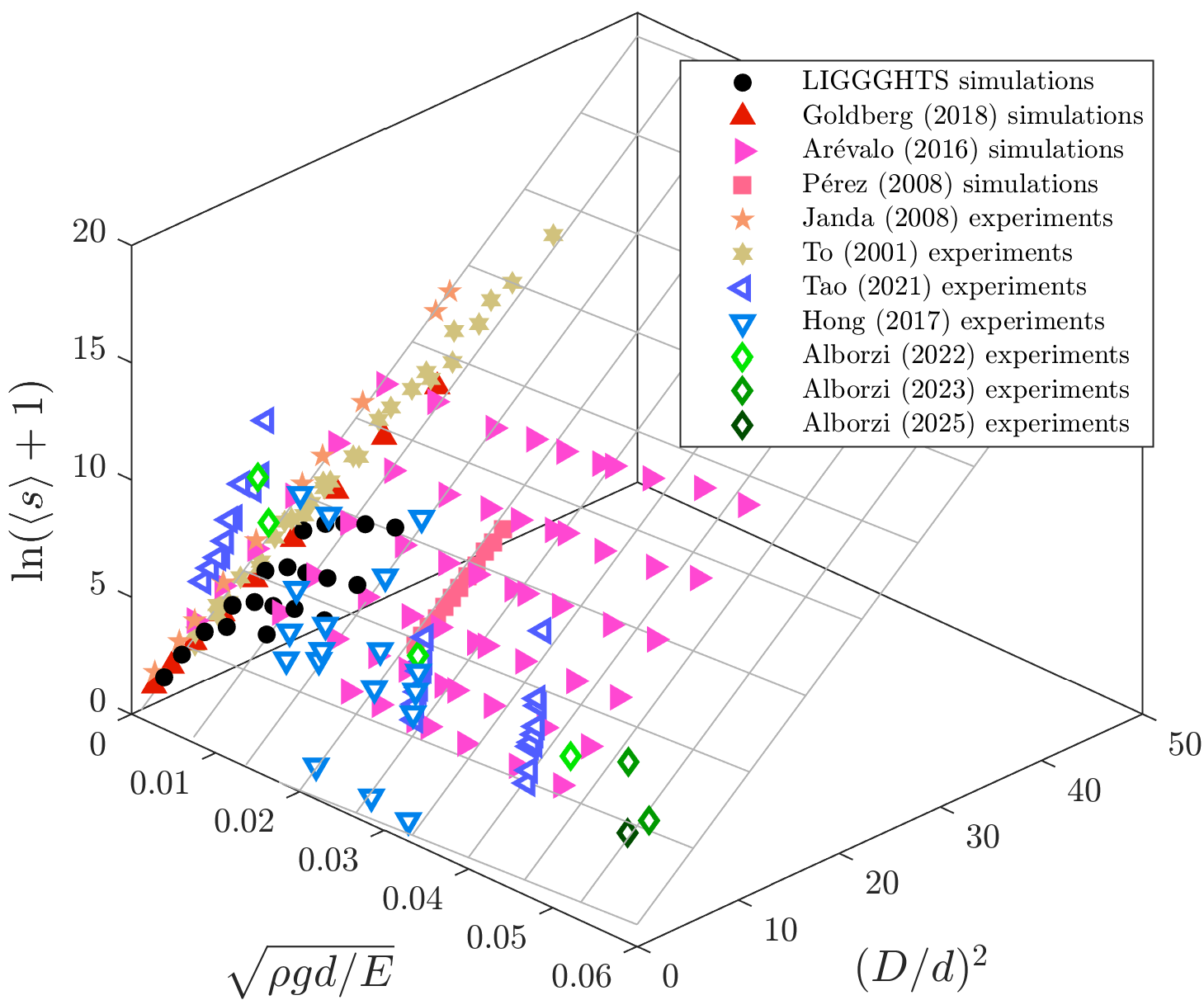}
    (b)\includegraphics[width=0.9\columnwidth]{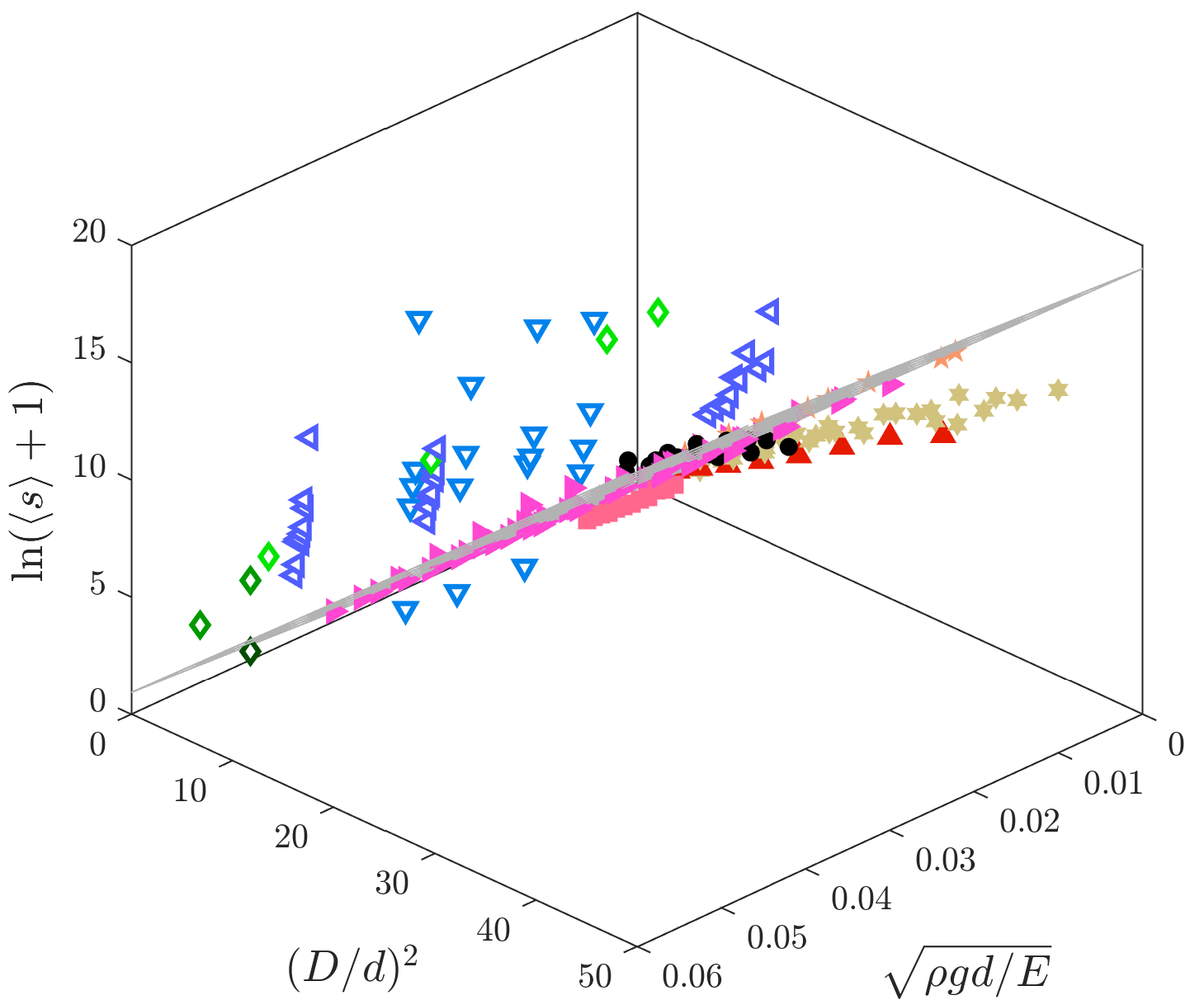}
    \caption{Mean avalanche size in a 2D silo as a function of the two dimensionless parameters. (a) and (b) show two different views of the 3D representation. Black dots represent our LIGGGHTS simulation data. Other symbols (see legend) display data from other sources: Refs.~\cite{Goldberg2018,Janda2008,To2001,Arevalo2016,Perez2008,Alborzi2022,Alborzi2023,Alborzi2025,Tao_PRE_2021,Hong_PRE_2017}. The data from Goldberg et al.~\cite{Goldberg2018} correspond to simulations in the hard particle limit. The plane fitting the data corresponds to $\ln(\s+1)=0.388 (\Pi_2-1)^2 + 22 \sqrt{\Pi_3}$. A rotating animation of this plot, as well as a Matlab .fig file, is available in the Supplemental Materials \cite{SM}.}
    
    \label{fig:2D}
\end{figure}

In addition to our simulations, we compiled available data for 2D systems to test our proposed scalings~\cite{Goldberg2018,Janda2008,To2001,Arevalo2016,Perez2008,Alborzi2022,Alborzi2023,Alborzi2025,Tao_PRE_2021,Hong_PRE_2017}.  The Young modulus and density for each experimental work was estimated based on the materials reported in each paper (see  Table \ref{tab:materials}). For simulations, estimating $E$ depends on the model used. The simulations of Goldberg et al. \cite{Goldberg2018} are based on a model of perfectly rigid particles, so we set $\Pi_3=0$. P\'erez \cite{Perez2008} uses a Hertz-3/2 contact model, so the estimation of $E$ in this case is straightforward. Ar\'evalo et al. \cite{Arevalo2016}, in contrast, model a linear spring contact interaction instead of the Hertzian model. Consistent with cylindrical particles, we had to estimate the thickness of the cylindrical particles to be equal to their diameter to be able to calculate the effective Young modulus used for the particles in this simulation. Finally, it is important to mention here that in several studies on clogging the mean avalanche size is not directly reported \cite{To2001,Alborzi2022,Alborzi2023}. In these works, clogging is characterized via the clogging probability $J(N)$, which is defined as the probability that the silo will clog before $N$ particles pass through the opening. However, we can estimate $\s$ form $J(N)$ as described in detail in Appendix A.

Figure~\ref{fig:2D} presents the data from our simulations (black points) along with the compiled results from previous authors for frictional grains~\cite{Goldberg2018,Janda2008,To2001,Arevalo2016,Perez2008} (closed symbols with red, pink, orange, and yellow colors) as well as clearly distinct soft, frictionless particle experiments for comparison~\cite{Alborzi2022,Alborzi2023,Alborzi2025,Tao_PRE_2021,Hong_PRE_2017} (open symbols of varying blue and green colors). All data from frictional particles is fairly consistent with Eq.~(\ref{eq:scaling}) and with our simulation data. Nevertheless, the results from To et al. \cite{To2001} and Goldberg et al. \cite{Goldberg2018} display somewhat lower mean avalanche sizes. We assert that this is likely due to differences in the actual protocol used to break clogging arches in these works.  Hence, we have fit  Eq.~(\ref{eq:scaling}) only to our LIGGGHTS simulation data (black points) and the data from Arévalo et al. \cite{Arevalo2016} (pink triangles), P\'erez \cite{Perez2008} (orange squares), and Janda et al. \cite{Janda2008} (orange stars). The fit to plane described by Eq.~(\ref{eq:scaling}) yields  $A_3=0.388 \pm 0.005$ and $B_3 = 22 \pm 3$. 
(The soft, frictionless particle experiments clearly do not align with the frictional particle studies or our simulations; we will address this shortly in Sec.~\ref{sec:soft}.)

Given the importance of dimensionality in the exponent of $D/d$ one cannot expect $A_2$ to agree with $A_3$. However, it is interesting to compare $B_2$ and $B_3$ since the square root scaling for $\rho g d/E$ is valid in 2D and 3D. Interestingly, we observe that $B_3 \approx B_2$ considering their uncertainties. 

\subsection{Soft particles in experiments\label{sec:soft}}

In quasi-2D configurations, clogging of soft, frictionless particles was reported by few research groups \cite{Hong_PRE_2017,Tao_PRE_2021,Alborzi2022,Alborzi2023,Alborzi2025}. Although these systems are not frictional, as the one we considered in our simulations, and most of these authors use hoppers instead of flat-bottom silos, we include the available data from these references in Fig.~\ref{fig:2D} (see open symbols of varying blue and green colors). In most cases, the values of $\s$ are not reported \cite{Hong_PRE_2017,Tao_PRE_2021,Alborzi2022,Alborzi2023}, so we convert clogging probability to $\s$ according the calculation we present in Appendix A. Most results from these experiments display mean avalanche sizes larger than those expected from the data with frictional particles and do not align with the fitted scaling law. Of course, larger avalanches are a natural consequence of the frictionless character of these particles. Surprisingly, a few data points for frictionless particles display smaller $\s$ that expected for a frictions system -- lie below the fitted plane. However, these points are obtained by converting clogging probability to $\s$. In such conversion, $\s$ presents large uncertainties when clogging probability is close to one, i.e., when the orifice size is close to the particle size $D/d \approx 1.0$.

It is worth mentioning that Wang et al. \cite{Wang_Soft_Matter_2021} and Harth et al. \cite{Harth2020} consider soft, frictionless particles flowing through an orifice in a quasi-2D, flat-bottom silo. However, these authors report transient clogging rather than stable arch formation. We therefore cannot include their data in the present work.

As we have mentioned above, a few works have considered 3D silos with soft particles \cite{Pongo2021,Stannarius2019}. Unfortunately, these particles are not only soft, but also frictionless. It seems this makes clogging a more transient phenomenon and permanent arches are less likely to form. As such, to our knowledge, there is not data on mean avalanche size or clogging probability available in the literature for 3D systems of soft particles. It would be interesting to see the development of clogging experiments using soft, frictional particles in the future. 

\section{Conclusions}

We have shown that dimensional analysis suggests that the mean avalanche size in a static, flat bottom silo with a small orifice can be put in terms of the orifice to particle size ratio and the ratio between the force of gravity on a particle and its stiffness. The available data in the literature complemented with our own DEM simulations of frictional soft particles are consistent with an exponential scaling that depends on dimensionality as given by Eq.~(\ref{eq:scaling}).

Our results help to put into a unified framework a number of previous studies on clogging in both 2D and 3D. Moreover, it suggests that new experiments using frictional soft particles, which are hard to fabricate, are of special interest to complete the full picture in this clogging phenomenon. Also, other variables that affect avalanche size can be included in the analysis such as hopper angle \cite{Lopez_Rodriguez_PRE_2019} and orifice inclination \cite{Sheldon2010}. These are already non-dimensional quantities that extend the parameter space. There is still little done in these directions. In relation to the friction coefficient, a non-dimensional quantity relating normal and tangential forces, the available data suggest that $\s$ is marginally affected and can be considered irrelevant for most purposes \cite{Zuriguel2005}. However, for very low friction coefficients ($\mu<0.1$) the mean avalanche size grows as the frictionless limit is approached \cite{Zablotsky2024}. This particular limit is of interest since most soft particles studied in the literature are indeed frictionless.   

Beyond material properties and silo geometry, there is yet a number of extensions that can be considered from this point. For example, clogging in vibrated systems are of particular interest. In this case, new variables come into consideration (vibration amplitude and frequency). However, such problems are not a simple extension of the current work because mean avalanche size is less relevant in vibrated systems. In such problems the exponent of the power-law tail of the distribution of clogging times is the measure of interest rather than avalanche sizes \cite{Janda2009,Mankoc2009}.

\begin{acknowledgments}
LAP acknowledges fruitful discussions with Iker Zuriguel, Angel Garcimartín and Diego Maza. LAP and RK thank Douglas Durian for valuable discussions. JM thanks CONICET for a scholarship. This work was funded by CONICET (Argentina) through grant PIP717 and Universidad Nacional de La Pampa (Argentina) through grant F65.
\end{acknowledgments}

\section*{Author Contributions}

J.M., L.A.P. and R.K. contributed equally to this work.

\bigskip

\appendix

\section{Calculating $\s$ from clogging probability}

When considering clogging, some authors report clogging probability $J(N)$, defined as the probability that the silo will clog before $N$ particles pass through the opening. This quantity can be used to estimate the mean avalanche size $\s$ if one assumes that the probability $p$ that a particle that arrives at the orifice will pass without clogging is constant. Under this assumption, the probability  $n(s)$ that exactly $s$ particles pass before a clog occurs is \cite{Zuriguel2005}
\begin{equation}\label{eq:ns}
    n(s) = p^s (1 - p).
\end{equation}
This avalanche size distribution is consistent with the exponential distribution observed by many authors \cite{To2005,Zuriguel2003,Thomas2015,Janda2008}. The mean of the distribution (\ref{eq:ns}) is
\begin{equation}\label{eq:s}
    \s = p/(1-p).
\end{equation}

We can also calculate $J(N)$ form $n(s)$. By definition,
\begin{equation}\label{eq:J1}
J(N) = 1 - \sum_{s=N}^{\infty} n(s).
\end{equation}
Plugging in (\ref{eq:ns}) into (\ref{eq:J1}), we obtain
\begin{align}\label{eq:J2}
J(N) &= 1 - \sum_{s=N}^{\infty} p^s (1 - p) = 1 - p^N.
\end{align}

Form Eq.~(\ref{eq:s}) and (\ref{eq:J2}) we can eliminate $p$ to obtain \cite{Janda2008}
\begin{align}
J(N) = 1 - p^N &= 1 - \left( \frac{\s}{1 + \s} \right)^N. 
\end{align}
Hence,
\begin{align}\label{eq:s2}
\s &= \frac{(1 - J(N))^{(1/N)}}{1 - (1 - J(N))^{(1/N)}}.
\end{align}

Equation (\ref{eq:s2}) allows us to convert the jamming probabilities $J$ at any system size $N$ to an estimate of the mean avalanche size $\s$. This is an important detail since different authors set their experiments (or simulations) with different system sizes, and this affects the clogging probability significantly \cite{Zuriguel2005}. Unfortunately, when one calculates $\s$ from Eq.~(\ref{eq:s2}), large uncertainties occur if $J$ is close to limiting values zero and one. This occurs when the orifice is too small ($D/d<2.0$) or too large ($D/d>4$).


\begin{thebibliography}{50}

\bibitem{Duran2000} J. Duran, Sands, Powders, and Grains: An Introduction to the Physics of Granular Materials, {\it Springer, New York} (2000).

\bibitem{Nedderman1992} R. M. Nedderman, Statics and Kinematics of Granular Materials, {\it Cambridge University Press, Cambridge} (1992).

\bibitem{Zuriguel2014} I. Zurigel, D. R Parisi, R. Cruz, C. Lozano, A. Janda, L. A. Pugnaloni, E. Cl{\'e}ment, Clogging transition of many-particle systems flowing through bottlenecks, {\it Sci. Rep.} {\bf 4}, 7324 (2014).

\bibitem{Zuriguel_PIP_2014} I. Zuriguel, Invited review: Clogging of granular materials in bottlenecks, {\it Pap. Phys.} {\bf 6}, 14  (2014).

\bibitem{Lopez_Rodriguez_PRE_2019} D. López-Rodríguez, D. Gella, K. To, D. Maza, A. Garcimartin, I. Zuriguel, Effect of hopper angle on granular clogging, {\it Phys. Rev. E} \textbf{99}, 032901 (2019).

\bibitem{To} K. To and H.-T. Tai, Flow and clog in a silo with oscillating exit, {\it Phys. Rev. E} \textbf{96}, 032906 (2017).

\bibitem{Parisi2023} Parisi, D. R., Wiebke, L. E., Mandl, J. N., Textor, J. (2023). Flow rate resonance of actively deforming particles, {\it Sci. Rep.} {\bf 13}(1), 9455.

\bibitem{Gharat_PowTech_2023} S. H. Gharat, L. A. Pugnaloni, Augmented flow and reduced clogging of particles passing through small apertures by addition of fine grains, {\it Pow. Tech.} \textbf{427}. 118695 (2023).

\bibitem{Madrid2021} M. A. Madrid, C. M. Carlevaro, L. A. Pugnaloni, M. Kuperman, S. Bouzat, Enhancement of the flow of vibrated grains through narrow apertures by addition of small particles,  \textit{Phys. Rev. E} \textbf{103}, L030901 (2021).

\bibitem{Hong_PRE_2017} Hong, X., Kohne, M., Morrell, M., Wang, H., Weeks, E. R. (2017). Clogging of soft particles in two-dimensional hoppers, {\it Phys. Rev. E} {\bf 96}(6), 062605.

\bibitem{Tao_PRE_2021} R. Tao, M. Wilson, E. R. Weeks, Soft particle clogging in two-dimensional hoppers, {\it Phys. Rev. E} {\bf 104}, 044909 (2021).

\bibitem{Wang_Soft_Matter_2021} J. Wang, B. Fan, T. Pongó, K. Harth, T. Trittel, R. Stannarius, M. Illig, T. Börzsönyi, R. C. Hidalgo, Silo discharge of mixtures of soft and rigid grains, {\it Soft Matter} \textbf{17}, 4282 (2021).

\bibitem{Alborzi2022} S. Alborzi, B. G. Clarka, S. M. Hashmi, Soft particles facilitate flow of rigid particles in a 2D hopper, {\it Soft Matter} {\bf 18}, 4127 (2022).

\bibitem{Alborzi2025} S. Alborzi, S. M. Hashmi, Low-friction soft particles add intermittency and avalanches to the flow of rigid particles in a hopper, {\it Powder Tech.} {\bf 459}, 121000 (2025).

\bibitem{Alborzi2023} S. Alborzi, D. Abrahamyan, S. M. Hashmi, Mixing particle stiffness in a two-dimensional hopper: Particle rigidity and friction enable variable arch geometry to cause clogging, {\it Phys. Rev. E} {\bf 107}, 024901 (2023).

\bibitem{Buckingham1914} E. Buckingham, On Physically Similar Systems; Illustrations of the Use of Dimensional Equations, {\it Phys. Rev.} {\bf 4}, 345 (1914).

\bibitem{Vaschy1892} A. Vaschy, Sur les lois de similitude en physique, {\it Annales t\'el\'egraphiques} {\bf 19}, 25 (1892).

\bibitem{Bertrand1879} J. Bertrand. Sur l’homogénéité dans les formules de Physique. {\it Comptes Rendus} {\bf 86}, 916 (1878).

\bibitem{Riabouchinsky1911} D. Riabouchinsky. Méthode des variables de dimension zéro et son application en aérodynamique, {\it L’Aérophile} {\bf 19}, 407 (1911).

\bibitem{To2005} K. To, Jamming transition in two-dimensional hoppers and silos, {\it Phys. Rev. E} {\bf 71}, 060301(R)(2005).

\bibitem{Janda2008} A. Janda, I. Zuriguel, A. Garcimartıin, L. A. Pugnaloni, D. Maza, Jamming and critical outlet size in the discharge of a two-dimensional silo, {\it Europhys. Lett.} {\bf 84}, 44002 (2008).

\bibitem{Thomas2015} C. C. Thomas, D. J. Durian, Fraction of clogging configurations sampled by granular hopper flow,  {\it Phys. Rev. Lett.} {\bf 114}, 178001 (2015).

\bibitem{Zuriguel2003} I. Zuriguel, L. A. Pugnaloni, A. Garcimartin, D. Maza, Jamming during the discharge of grains from a silo described as a percolating transition, {\it Phys. Rev. E} {\bf 68}, 030301(R) (2003).

\bibitem{Zuriguel2005} I. Zuriguel, A. Garcimartín, D. Maza, L. A. Pugnaloni, J. M. Pastor, Jamming during the discharge of granular matter from a silo, {\it Phys. Rev. E} {\bf 71}, 051303 (2005).

\bibitem{Arevalo2016} R. Arévalo, I. Zuriguel, Clogging of granular materials in silos: effect of gravity and outlet size, {\it Soft Matter} {\bf 12}(1) 123 (2016).

\bibitem{To2001} K. To, P. Y. Lai, H. K. Pak, Jamming of granular flow in a two-dimensional hopper {\it Phys. Rev. Lett.} {\bf 86}, 71 (2001).

\bibitem{Hong2022} X. Hong, K. W. Desmond, D. Chen, E. R. Weeks, Clogging and avalanches in quasi-two-dimensional emulsion hopper flow, {\it Phys. Rev. E} {\bf 105}, 014603 (2022). 

\bibitem{Zablotsky2024} A. Zablotsky, M. A. Madrid, C. M. Carlevaro, M. Kuperman, L. A. Pugnaloni, S. Bouzat, Reduction of clogging of vibrated grains passing through a narrow aperture by the addition of low-friction particles {\it Phys. Rev. E} {\bf 110}, 034902 (2024).

\bibitem{Kloss2012} C. Kloss, C. Goniva, A. Hager, S. Amberger, S. Pirker, Models, algorithms and validation for opensource DEM and CFD–DEM, {\it Prog. Comput. Fluid Dyn.} {\bf 12}, 140 (2012).

\bibitem{SM} See Supplemental Material at [URL will be inserted by publisher] for additional files for more clear visualization and interactivity of Figures 3 and 4..


\bibitem{Goldberg2018} E. Goldberg, C. M. Carlevaro, L. A. Pugnaloni, Clogging in two-dimensions: effect of particle shape, {\it J. Stat. Mech. Theo. Exp.} {\bf 11}, 113201 (2018).

\bibitem{Stannarius2019} R. Stannarius, D. Sancho Martinez. T. Finger, E. Somfai, T. Börzsönyi, Packing and flow profiles of soft grains in 3D silos reconstructed with X‐ray computed tomography, {\it Granular Matter} {\bf 21}, 56 (2019).

\bibitem{Pongo2021} T. Pongó, R. Stannarius, V. Stiga, J. Török, S. Lévay, B. Szabó, R. Cruz Hidalgo, T. Börzsöny, Flow in an hourglass: particle friction and stiffness matter, {\it New J. Phys.} {\bf 23}, 023001 (2021).

\bibitem{Perez2008} G. P\'{e}rez, Numerical simulations in granular matter: The discharge of a 2D silo, {\it Pramana J. Phys.} {\bf 70}, 6 (2008).


\bibitem{Harth2020} K. Harth, J. Wang, T. Börzsönyi, Ralf Stannarius, Intermittent flow and transient congestions of soft spheres passing narrow orifices, {\it Soft Matter} {\bf 16}, 8013 (2020).

\bibitem{Sheldon2010} H. G. Sheldon, D. J. Durian, Granular discharge and clogging for tilted hoppers,
{\it Granular Matter} {\bf 12}, 579 (2010).

\bibitem{Mankoc2009} C. Mankoc, A. Garcimartín, I. Zuriguel, D. Maza, and L. A. Pugnaloni, Role of vibrations in the jamming and unjamming of grains discharging from a silo {\it Phys. Rev. E} \textbf{80}, 011309 (2009).

\bibitem{Janda2009} A. Janda, D. Maza, A. Garcimart\'in, E. Kolb, J. Lanuza, E. Cl\'ement, Unjamming a granular hopper by vibration, {\it Europhys. Lett.} {\bf 87}, 24002 (2009).

\end{thebibliography}
\end{document}